\documentstyle[aps,preprint,epsf,12pt]{revtex}

\draft
\title{QUANTUM EFFECTS AND CLUSTER FORMATION}
\author{ALI SHOJAI\thanks{Email: SHOJAI@THEORY.IPM.AC.IR}}
\address{Physics Department, Tarbiat Modares University,P.O.Box 14155--4838, Tehran, IRAN}
\address{and}
\address{Institute for Studies in Theoretical Physics and Mathematics, P.O.Box 19395-5531, Tehran,
IRAN}
\author{FATIMAH SHOJAI\thanks{Email: FATIMAH@THEORY.IPM.AC.IR}}
\address{Physics Department, Iran University of Science and Technology, P.O.Box 16765--163, Narmak, Tehran,
IRAN}
\address{and}
\address{Institute for Studies in Theoretical Physics and Mathematics, P.O.Box 19395-5531, Tehran,
IRAN}
\begin{document}
\maketitle
\begin{abstract}
The causal interpretation of quantum mechanics is applied to the
universe as a whole and the problem of cluster formation is
studied in this framework. It is shown that the quantum effects
{\it may\/} be the source of the cluster formation.
\end{abstract}
\section{CAUSAL QUANTUM MECHANICS AND GRAVITY}
Since the early years of 20th century, there is a causal theory of
quantum phenomena called {\it de-Broglie--Bohm\/}
theory\cite{de1,de2,Bohm,Hiley,Holland}. It is well proved that
the causal theory reproduces all the results of the orthodox
quantum theory\cite{Hiley,Holland}, as well as predicting some new
results (such as time of tunneling through a
barrier\cite{Cushing}) which in principle lets the experiment to
choose between the orthodox and the causal quantum theories.
Perhaps the most important point about the causal theory is that
it presents a causal deterministic description of the reality. But
an important by product of this theory is that enables one to put
a step towards constructing a successful quantum gravity
theory\cite{GEO,CON,STQ,NLO,QGG,NMS}.

The causal quantum theory is based on two
postulates\cite{Bohm,Hiley,Holland}:
\begin{description}
\item[{\bf First Law:}] The physical reality is described by
$\{\vec{r}(t);\Psi(\vec{x},t)\}$ where $\vec{r}(t)$ is the
position vector of the particle and $\Psi(\vec{x},t)$ is a
self--field. Both have a causal evolution. $\Psi(\vec{x},t)$
satisfies an appropriate wave equation (like the Schr\"odinger
equation for a non-relativistic particle) while $\vec{r}(t)$
satisfies Newton's equation of motion in which there is an
additional force called {\it quantum force\/} resulted from
$\Psi(\vec{x},t)$. This force can be derived from some {\it
quantum potential\/} given by ${\cal
Q}=-\frac{\hbar^2}{2m}\frac{\nabla^2|\Psi|}{|\Psi|}$ for a non
relativistic particle.
\item[{\bf Second Law:}] The statistical features of the theory is
such that the ensemble density is given by $\rho=|\Psi|^2$.
\end{description}
So the complete description of the physical reality is given by
the following two equations for non--relativistic particles:
\begin{equation}
i\hbar\frac{\partial \Psi}{\partial t}=-\frac{\hbar^2}{2m}\nabla^2
\Psi+V\Psi
\end{equation}
\begin{equation}
m\frac{d^2\vec{r}}{dt^2}=-\left .\vec{\nabla}\left (V+{\cal
Q}\right )\right |_{\vec{x}=\vec{r}(t)};\ \ \ \ \ \ {\cal Q}=
-\frac{\hbar^2}{2m}\frac{\nabla^2|\Psi|}{|\Psi|}
\end{equation}

The relativistic extension is easy\cite{Hiley,Holland}. If one
works in the flat Minkowski space--time one should replace the
Schr\"odinger equation with some relativistic wave equation (such
as the Klein--Gordon equation for spin zero particles). The
quantum potential would be generalized to:
\begin{equation}
{\cal M}^2=m^2(1+{\cal Q})=m^2+\alpha\frac{\Box|\Psi|}{|\Psi|};\ \
\ \ \alpha=\frac{\hbar^2}{m^2c^2}
\end{equation}
and the equation of motion is given by:
\begin{equation}
{\cal M}\frac{dx^\mu}{d\tau^2}=\left (
\eta^{\mu\nu}-\frac{dx^\mu}{d\tau}\frac{dx^\nu}{d\tau}\right )
\partial_\nu {\cal M}
\end{equation}
It is important to note that ${\cal M}$ plays the role of the
(quantum) mass field of the
particle\cite{Holland,GEO,CON,STQ,QGG,NMS}.

The extension to the curved space--time is
trivial\cite{Holland,GEO,CON,STQ,QGG,NMS}:
\begin{equation}
{\cal M}^2=m^2+\alpha\frac{\Box|\Psi|}{|\Psi|};\ \ \ \
\alpha=\frac{\hbar^2}{m^2c^2}
\end{equation}
\begin{equation}
{\cal M}\frac{dx^\mu}{d\tau^2}=\left (
g^{\mu\nu}-\frac{dx^\mu}{d\tau}\frac{dx^\nu}{d\tau}\right )
\nabla_\nu {\cal M}\label{hehe}
\end{equation}

An important result of all these things is that the quantum
effects of matter are in fact geometrical in nature\cite{GEO}. One
can easily transforms the above equation of motion to the ordinary
geodesic equation via a conformal transformation of the
form\cite{GEO,CON,STQ,QGG}:
\begin{equation}
g_{\mu\nu} \rightarrow \widetilde{g}_{\mu\nu}=\frac{{\cal
M}^2}{m^2} g_{\mu\nu} \label{z}
\end{equation}
So that the matter quantum effects are included in the conformal
degree of freedom of the space--time metric\cite{GEO,QGG}. A
corollary of this result is that one can always work in a gauge
(classic gauge) in which no quantum effect be present or in a
gauge (quantum gauge) in which the conformal degree of freedom of
the space--time metric is identified with the quantum effects of
matter\cite{QGG}.

An important question is that if this quantum gravity theory leads
to some new results. This theory is investigated for the BigBang
and black holes in reference \cite{GEO,QGG}. But here we are
interested in investigating whether this theory has anything to do
with the cluster formation or clustering of the initial uniform
distribution of matter in the universe. The problem of cluster
formation is an important problem of cosmology and there are
several ways to tackle with it\cite{gal}. Here we don't want to
discuss those theories, and our claim is not that the present work
is a good one. Here we only state that {\it the cluster formation
can also be understood in this way\/}. It is a further task to
decide if this work is in complete agreement with experiment or
not.
\section{CLUSTER FORMATION AND THE QUANTUM EFFECTS}
In order to investigate if the cluster formation can be viewed as
a quantum effect produced by the quantum potential, we consider
the universe as a fluid. The hydrodynamics equation is given by:
\begin{equation}
{\cal M}\left (\frac{\partial p}{\partial
x^\nu}g^{\mu\nu}+\frac{1}{\sqrt{-g}}\left ( \sqrt{-g}(p+\rho)U^\mu
U^\nu \right ) +\Gamma^\mu_{\nu\lambda}(p+\rho)U^\mu U^\nu\right
)= \rho \left ( g^{\mu\nu}-U^\mu U^\nu \right )
\frac{\partial{\cal M}}{\partial x^\nu} \label{a}
\end{equation}
where we have introduced the quantum force in the right hand side
just as it is introduced in the equation of motion of a single
particle (see equation (\ref{hehe})). It must be noted that the
metric itself must be calculated from the corrected Einstein's
equations including the back--reaction terms\cite{GEO,QGG}. In
fact one must solve the above equation and the metric equation
simultaneously to obtain the metric and the density. We shall not
do in this way because solving those equations (equations of
references \cite{GEO,QGG}) is difficult. We shall do in a similar
way. It is an iterative way and is based on the fundamental
assumption of this theory, that is equation (\ref{z}). As the
first order of iteration, we consider the space--time metric as
given by the classical Einstein's equations (Robertson--Walker
metric) and solve the above equation for the density, then using
the result obtained, calculate the quantum metric using equation
(\ref{z}). Then the new metric can be used to obtain the density
at the second order and so on.

In the comoving frame and with the assumption that the universe is
in the dust mode ($p=0$) with the flat Robertson--Walker metric,
we have from the equation (\ref{a}):
\begin{equation}
\frac{d\rho^{(1)}}{dt}+3H\rho^{(1)}=0 \label{b}
\end{equation}
\begin{equation}
\frac{\partial{\cal Q}^{(1)}}{\partial x^i}=0
\end{equation}
where $^{(1)}$ denotes the first order of iteration and
$H=\dot{a}/a$ is the Hubble's parameter. The solution of the above
two equations is:
\begin{equation}
\rho^{(1)}=\frac{{\cal X}^{(1)2}}{t^2};\ \ \ \ \ \ \ \ \ \ {\cal
Q}^{(1)}=\text{constant}
\end{equation}
where ${\cal X}^{(1)}$ should yet be determined. The constancy of
the quantum potential leads to:
\begin{equation}
{\cal
Q}^{(1)}=\frac{\alpha}{2}\frac{\Box\sqrt{\rho^{(1)}}}{\sqrt{\rho^{(1)}}}=
-\frac{\alpha}{2a^2} \frac{\nabla^2{\cal X}^{(1)}}{{\cal X}^{(1)}}
\end{equation}
so that:
\begin{equation}
\nabla^2 {\cal X}^{(1)}+\beta {\cal X}^{(1)}=0 \label{c}
\end{equation}
where:
\begin{equation}
\beta=\frac{2{\cal Q}^{(1)}a^2}{\alpha}
\end{equation}
This equation for ${\cal X}^{(1)}$ can simply be solved either in
the Cartesian coordinates or in the spherical ones. The solution
is:
\begin{equation}
{\cal X}^{(1)}=\sin\left ( \sqrt{\frac{\beta}{3}}x\right
)\sin\left ( \sqrt{\frac{\beta}{3}}y\right )\sin\left (
\sqrt{\frac{\beta}{3}}z\right ) \label{d}
\end{equation}
or:
\begin{equation}
{\cal X}^{(1)}=\sum_{l,m}\left ( a_{lm}j_l(\sqrt{\beta}r)+
b_{lm}n_l(\sqrt{\beta}r)\right ) Y_{lm}(\theta,\phi) \label{e}
\end{equation}
This is the first order approximation. At the second order, one
must use the equation (\ref{z}) to change the scale factor $a^2$
to $a^2(1+{\cal Q})$, and then from the relation (\ref{b}) we
have:
\begin{equation}
a^2(1+{\cal Q})=t^{2/3}{\cal X}^{(1)-4/3}
\end{equation}
So that:
\begin{equation}
{\cal Q}^{(2)}=-1+{\cal X}^{(1)-4/3}
\end{equation}
and then using this form of the quantum potential in the relation
(\ref{c}) or (\ref{d}) leads to the following approximation for
the density:
\begin{equation}
\rho^{(2)}=\frac{1}{t^2}\sin^2\left (
\sqrt{\frac{\gamma}{3}}x\right )\sin^2\left (
\sqrt{\frac{\gamma}{3}}y\right )\sin^2\left (
\sqrt{\frac{\gamma}{3}}z\right ) \label{f}
\end{equation}
where:
\begin{equation}
\gamma=\frac{2a^2}{\alpha}\left (-1+{\cal X}^{(1)-4/3}\right )
\end{equation}
and ${\cal X}^{(1)}$ is given by the relation (\ref{d}). This
procedure can be done order by order.
\section{RESULTS}
\begin{figure}
\vspace{1.2in}
\begin{center}
\epsffile{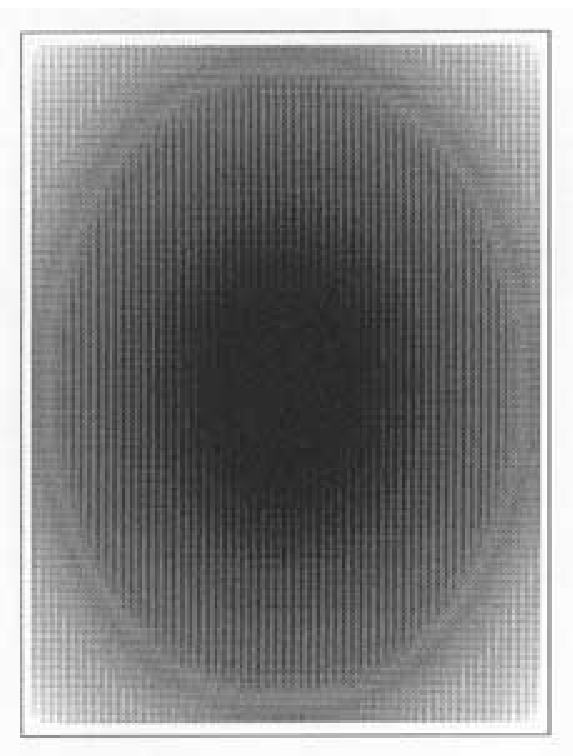}
\end{center}
\caption{Cartesian mode after expansion by the factor 1.1}
\label{1}
\end{figure}

\begin{figure}
\vspace{1.2in}
\begin{center}
\epsffile{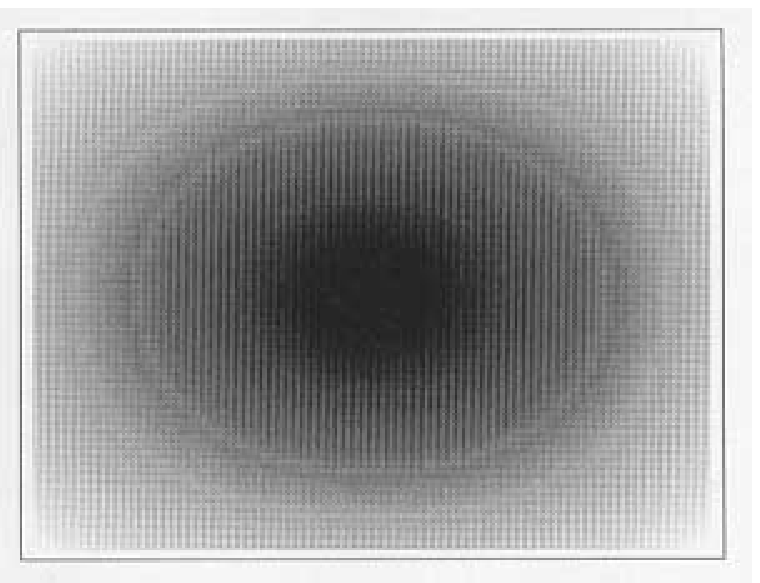}
\end{center}
\caption{Cartesian mode after expansion by the factor 1.2}
\label{5}
\end{figure}

\begin{figure}
\vspace{1.2in}
\begin{center}
\epsffile{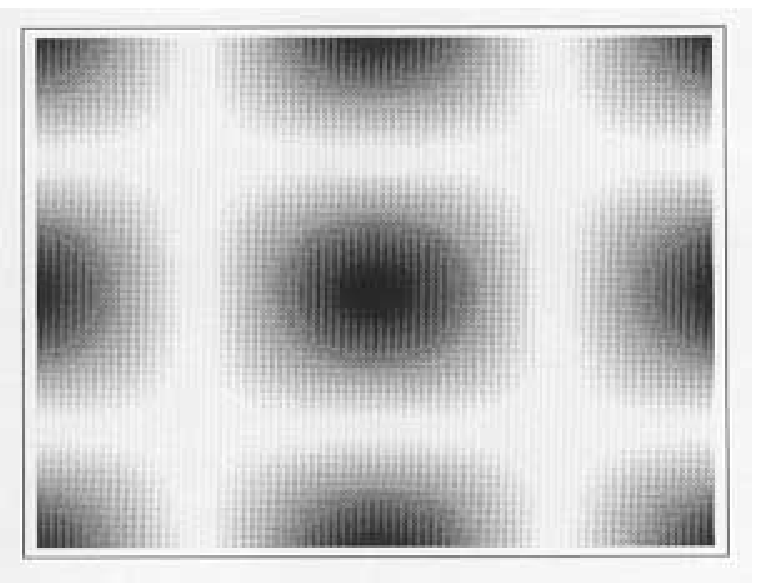}
\end{center}
\caption{Cartesian mode after expansion by the factor 3}
\label{10}
\end{figure}

\begin{figure}
\vspace{1.2in}
\begin{center}
\epsffile{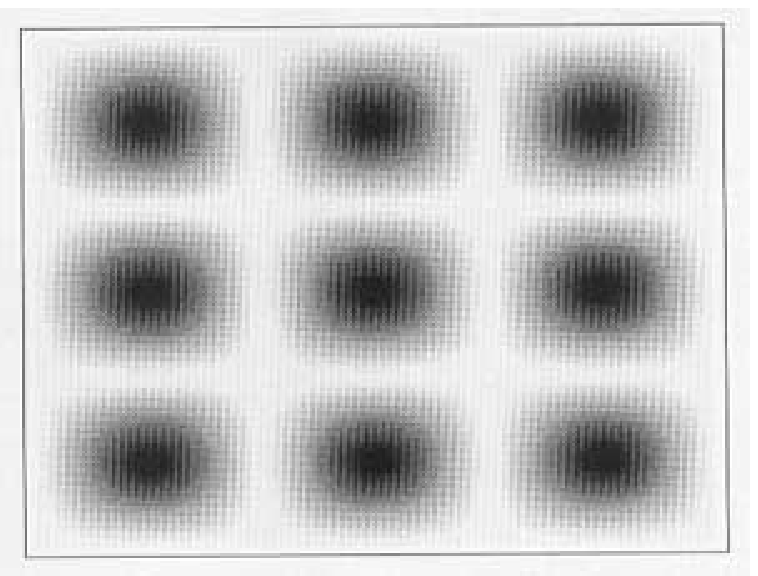}
\end{center}
\caption{Cartesian mode after expansion by the factor 5}
\label{15}
\end{figure}

\begin{figure}
\vspace{1.2in}
\begin{center}
\epsffile{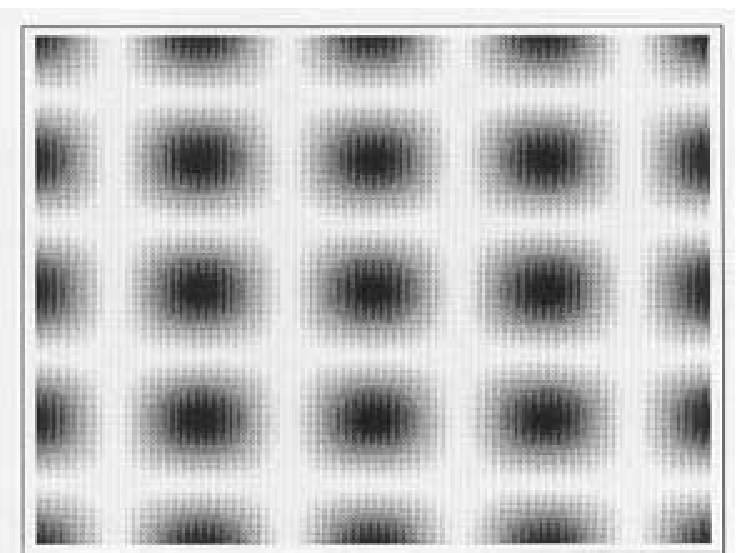}
\end{center}
\caption{Cartesian mode after expansion by the factor 6}
\label{20}
\end{figure}

In the figures (\ref{1}), (\ref{5}), (\ref{10}), (\ref{15}) and
(\ref{20}) the density at four times are shown and the clustering
can be seen easily. These figures are plotted using the solution
in the Cartesian coordinates.

\begin{figure}
\vspace{1.2in}
\begin{center}
\epsffile{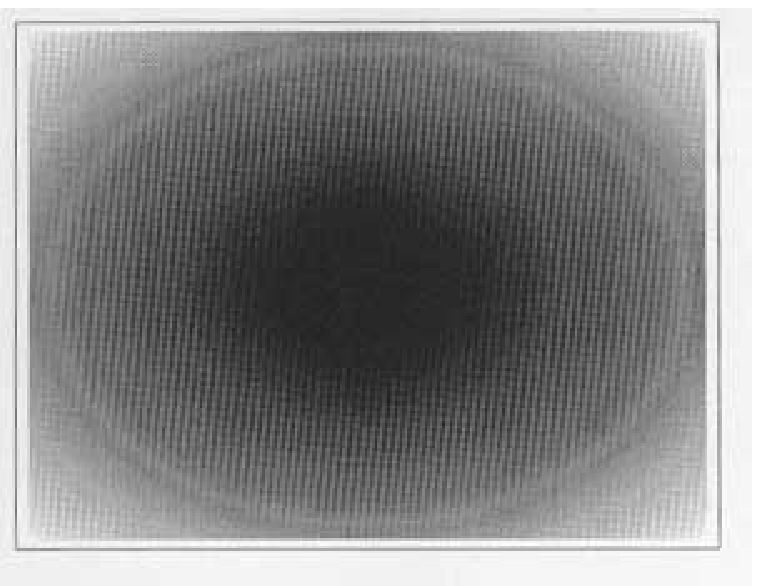}
\end{center}
\caption{$(00)$ mode after expansion by the factor 1.1}
\label{001}
\end{figure}

\begin{figure}
\vspace{1.2in}
\begin{center}
\epsffile{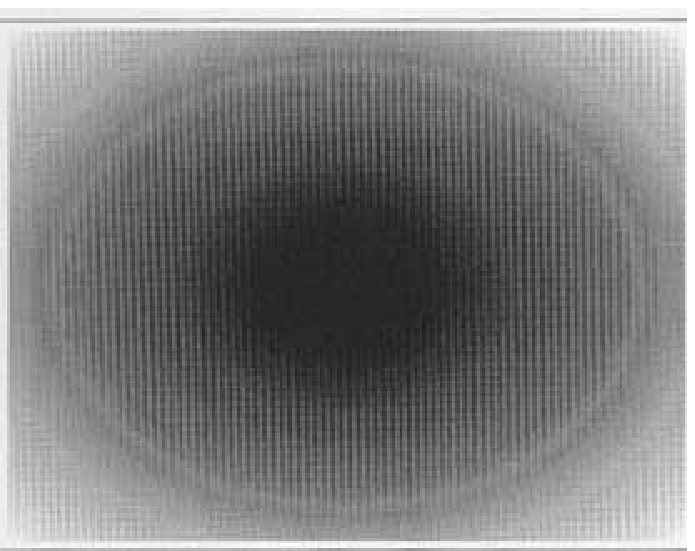}
\end{center}
\caption{$(00)$ mode after expansion by the factor 1.2}
\label{002}
\end{figure}

\begin{figure}
\vspace{1.2in}
\begin{center}
\epsffile{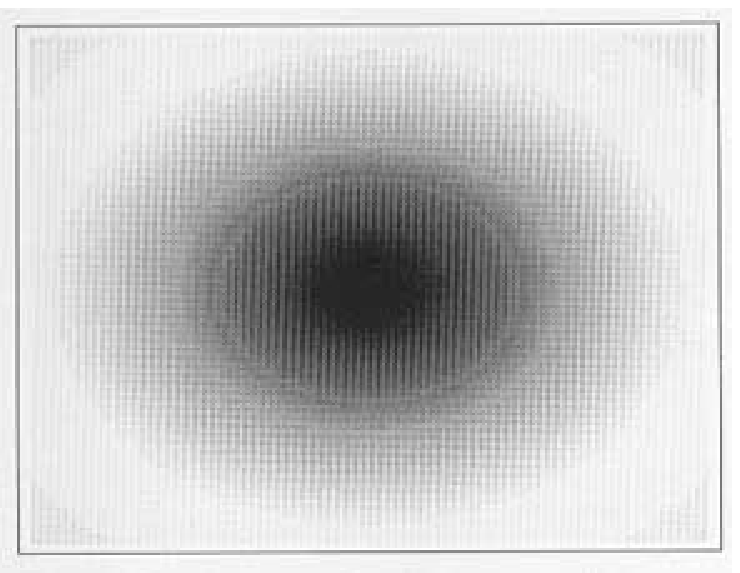}
\end{center}
\caption{$(00)$ mode after expansion by the factor 3} \label{003}
\end{figure}

\begin{figure}
\vspace{1.2in}
\begin{center}
\epsffile{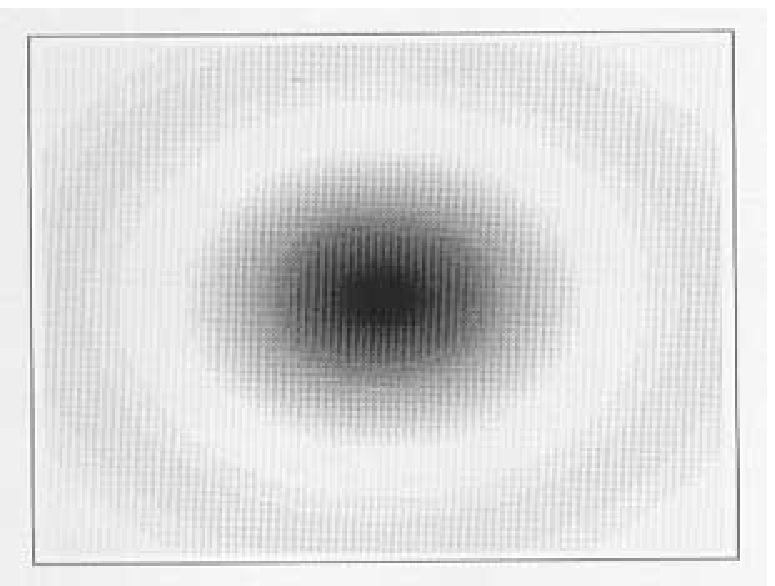}
\end{center}
\caption{$(00)$ mode after expansion by the factor 5} \label{004}
\end{figure}

\begin{figure}
\vspace{1.2in}
\begin{center}
\epsffile{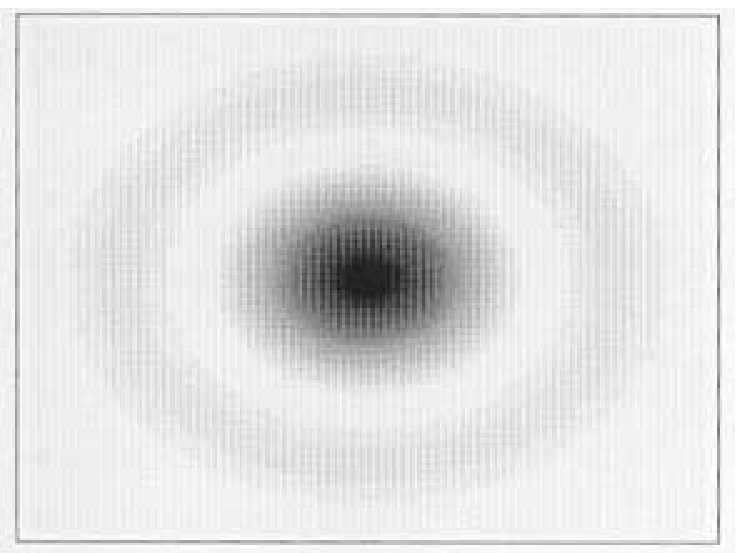}
\end{center}
\caption{$(00)$ mode after expansion by the factor 6} \label{005}
\end{figure}

In figures (\ref{001}), (\ref{002}), (\ref{003}), (\ref{004}) and
(\ref{005}) the $(l,m)=(00)$ mode of equation (\ref{e}) is shown
at five time steps.

\begin{figure}
\vspace{1.2in}
\begin{center}
\epsffile{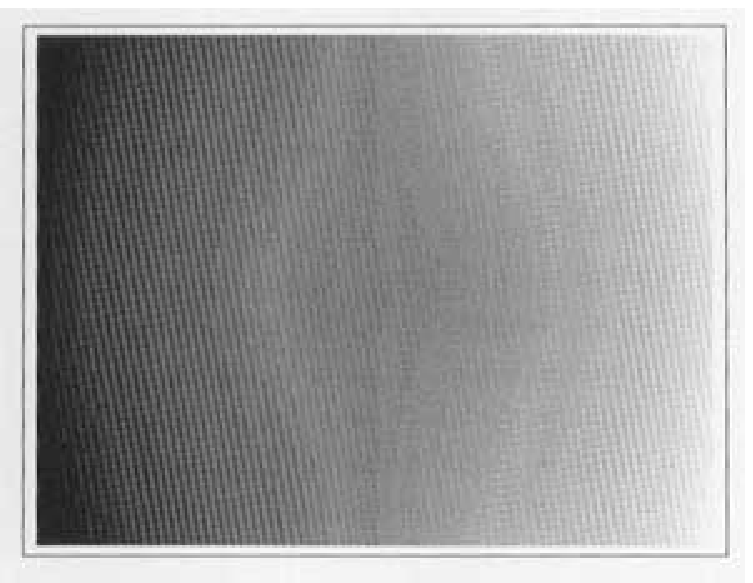}
\end{center}
\caption{$(11)\oplus (1-1)$ mode after expansion by the factor
1.2} \label{11}
\end{figure}

\begin{figure}
\vspace{1.2in}
\begin{center}
\epsffile{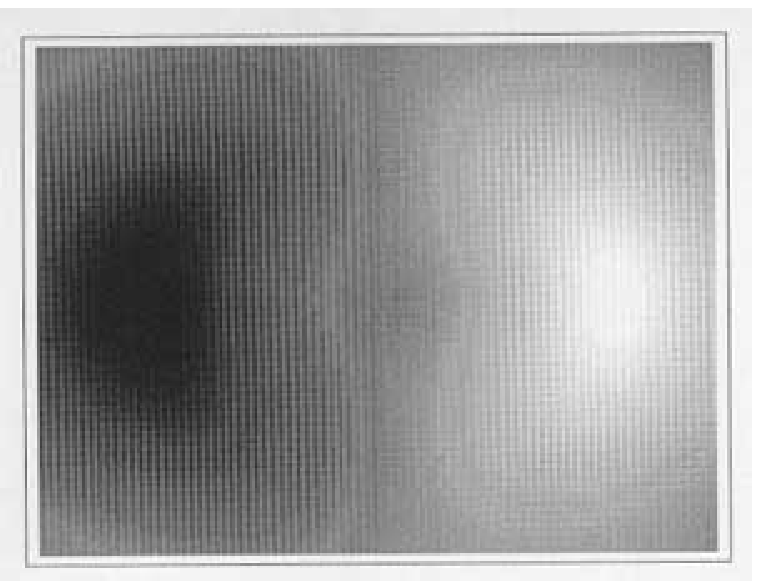}
\end{center}
\caption{$(11)\oplus (1-1)$ mode after expansion by the factor 3}
\label{12}
\end{figure}

\begin{figure}
\vspace{1.2in}
\begin{center}
\epsffile{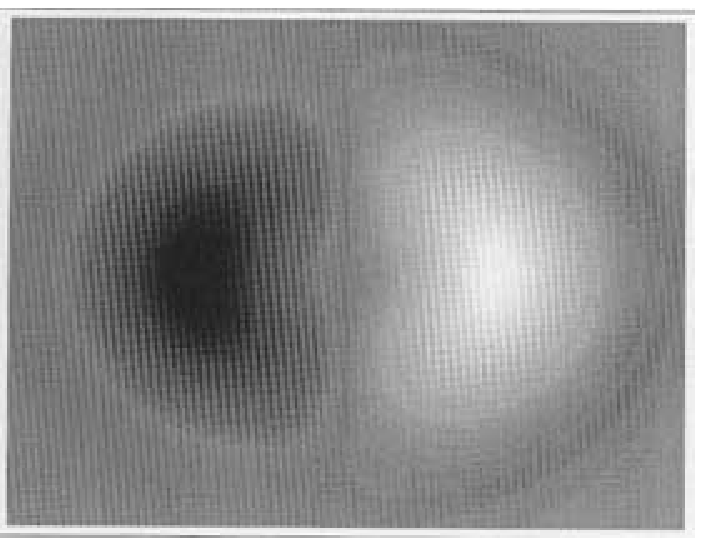}
\end{center}
\caption{$(11)\oplus (1-1)$ mode after expansion by the factor 5}
\label{13}
\end{figure}

\begin{figure}
\vspace{1.2in}
\begin{center}
\epsffile{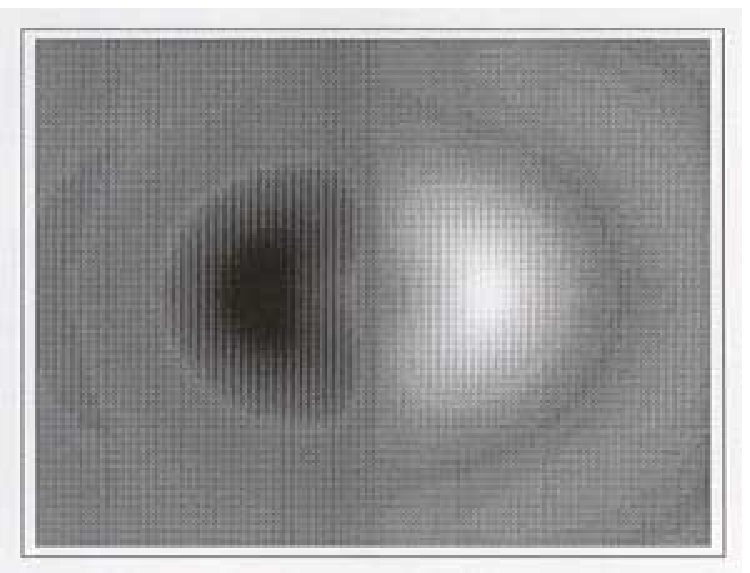}
\end{center}
\caption{$(11)\oplus (1-1)$ mode after expansion by the factor 6}
\label{14}
\end{figure}

\begin{figure}
\vspace{1.2in}
\begin{center}
\epsffile{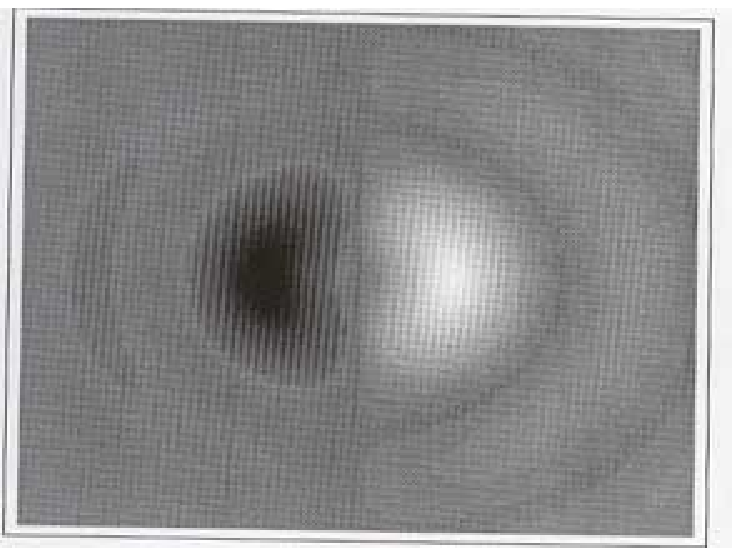}
\end{center}
\caption{$(11)\oplus (1-1)$ mode after expansion by the factor 8}
\label{15a}
\end{figure}

In figures (\ref{11}), (\ref{12}), (\ref{13}), (\ref{14}) and
(\ref{15a}) the $(11)\oplus (1-1)$ mode is shown at five time
steps.

\begin{figure}
\begin{center}
\epsffile{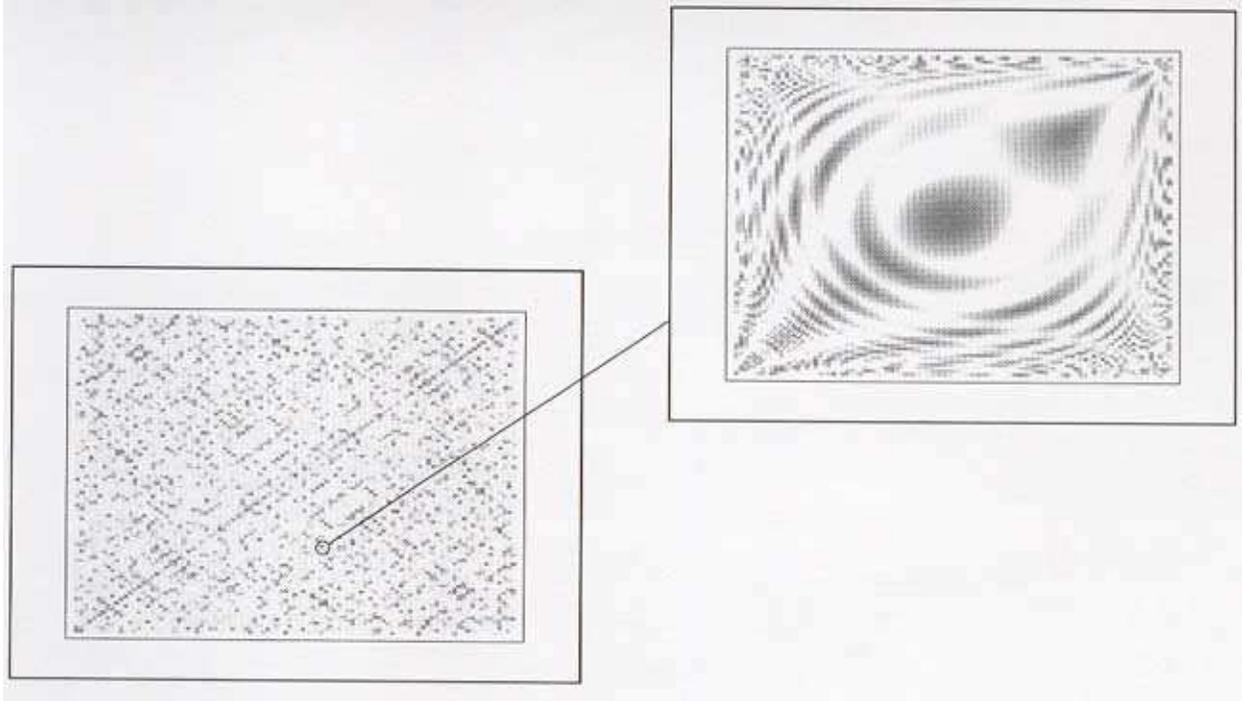}
\end{center}
\caption{Second order solution after expansion by the factor 8}
\label{xxx}
\end{figure}

The second order solution (\ref{f}) is shown in figure \ref{xxx},
where both large scale and small scale structures are shown.

It is important to note that the clustering can be seen in any of
these figures. In the last figure, however, one observes that at
the large scale the universe is homogeneous and isotropic, while
at the small scale these symmetries are broken.

\begin{figure}
\begin{center}
\epsffile{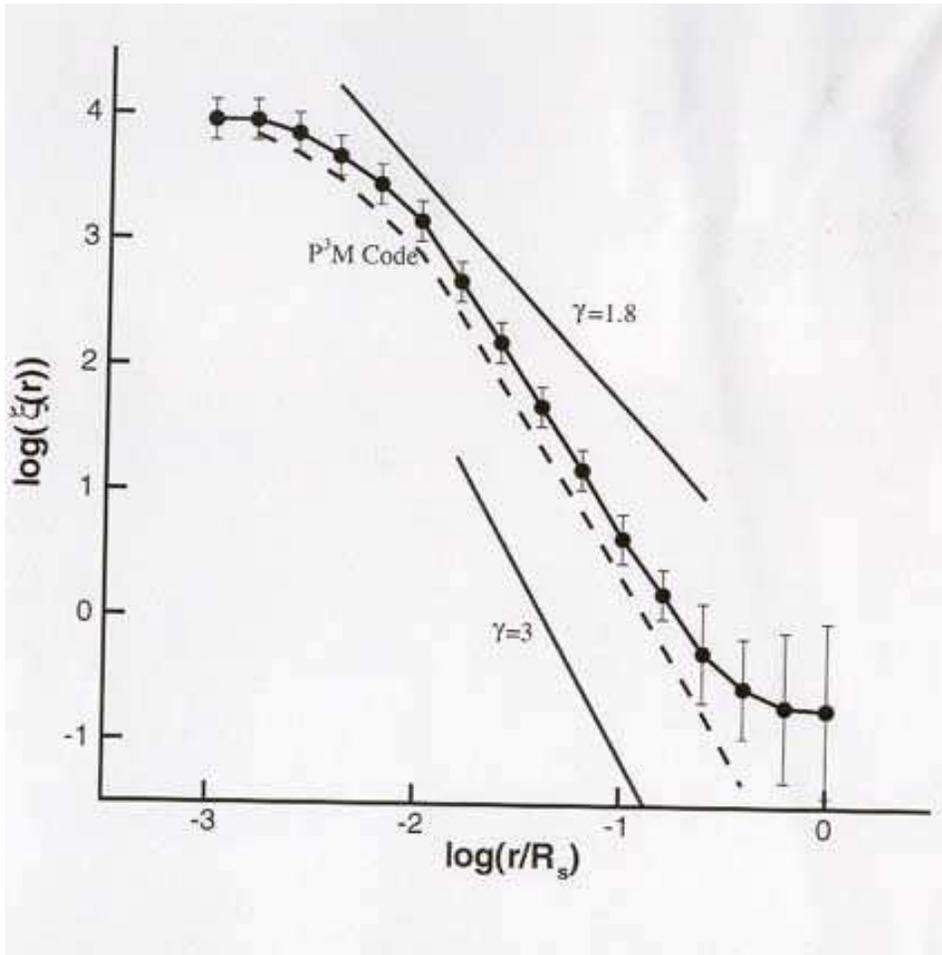}
\end{center}
\caption{Correlation function after expansion by the factor 8}
\label{ccc}
\end{figure}

At the end, in order to see whether or results are in agreement
with the observed clustering, the correlation function ($\xi(r)$)
is obtained from the  third order of iteration and is compared
with the cases $\xi=(r/r_0)^{-\gamma}$ with $\gamma=1.8$ and
$\gamma=3$ and with the standard result of a typical $P^3M$
code\cite{gal}. As it can be seen in figure (\ref{ccc}) our
results are in good agreement with the $P^3M$ code and with
observation.

As we stated previously, our claim here is not that this theory is
a good one for the cluster formation problem. But it is only
claimed that in the framework of causal quantum theory, the
quantum force {\it may\/} be a cause for the cluster formation.

\epsfxsize=3in \epsfysize=3in
\newpage$\rightarrow$
\epsfxsize=3in \epsfysize=3in
\newpage$\rightarrow$
\epsfxsize=3in \epsfysize=3in
\newpage$\rightarrow$
\epsfxsize=3in \epsfysize=3in
\newpage$\rightarrow$
\epsfxsize=3in \epsfysize=3in
\newpage$\rightarrow$
\epsfxsize=3in \epsfysize=3in
\newpage$\rightarrow$
\epsfxsize=3in \epsfysize=3in
\newpage$\rightarrow$
\epsfxsize=3in \epsfysize=3in
\newpage$\rightarrow$
\epsfxsize=3in \epsfysize=3in
\newpage$\rightarrow$
\epsfxsize=3in \epsfysize=3in
\newpage$\rightarrow$
\epsfxsize=3in \epsfysize=3in
\newpage$\rightarrow$
\epsfxsize=3in \epsfysize=3in
\newpage$\rightarrow$
\epsfxsize=3in \epsfysize=3in
\newpage$\rightarrow$
\epsfxsize=3in \epsfysize=3in
\newpage$\rightarrow$
\epsfxsize=3in \epsfysize=3in
\newpage$\rightarrow$
\epsfxsize=5in \epsfysize=5in
\end{document}